\documentclass[aps,pra,superscriptaddress,twocolumn,floatfix]{revtex4-2}
\usepackage{setspace}
\usepackage[T1]{fontenc}
\usepackage[latin9]{inputenc}
\usepackage{amstext}
\usepackage{amsmath}
\usepackage{amsfonts}
\usepackage{graphicx}
\usepackage{graphics}
\usepackage{subfigure}
\usepackage{esint}
\usepackage{color}
\usepackage[normalem]{ulem}
\usepackage[colorlinks, linkcolor=blue, citecolor=blue,hyperindex,bookmarks=false,pdfstartview=FitH]{hyperref}
\usepackage{lineno}
\makeatletter

\@ifundefined{textcolor}{}
{%
 \definecolor{BLACK}{gray}{0}
\definecolor{WHITE}{gray}{1}
 \definecolor{RED}{rgb}{1,0,0}
 \definecolor{GREEN}{rgb}{0,1,0}
 \definecolor{BLUE}{rgb}{0,0,1}
 \definecolor{CYAN}{cmyk}{1,0,0,0}
 \definecolor{MAGENTA}{cmyk}{0,1,0,0}
 \definecolor{YELLOW}{cmyk}{0,0,1,0}
}
\makeatother

\makeatother

\begin{document}

\title{Quantum Shortcut to Adiabaticity for State Preparation in a Finite-Sized Jaynes-Cummings Lattice}

\author{Kang Cai} 
\affiliation{School of Natural Sciences, University of California, Merced, California 95343, USA}

\author{Prabin Parajuli}
\affiliation{School of Natural Sciences, University of California, Merced, California 95343, USA}

\author{Anuvetha Govindarajan}
\affiliation{School of Natural Sciences, University of California, Merced, California 95343, USA}

\author{Lin Tian}
\email{ltian@ucmerced.edu}
\affiliation{School of Natural Sciences, University of California, Merced, California 95343, USA}

\begin{abstract}
In noisy quantum systems, achieving high-fidelity state preparation using the adiabatic approach faces a dilemma: either extending the evolution time to reduce diabatic transitions or shortening it to mitigate decoherence effects.
Here, we present a quantum shortcut to adiabaticity for state preparation in a finite-sized Jaynes-Cummings lattice by applying counter-diabatic (CD) driving along given adiabatic trajectories. Leveraging the symmetry of eigenstates in our system, we convert the CD driving to an implementable Hamiltonian that only involves local qubit-cavity couplings for a two-site lattice with one polariton excitation. Additionally, we derive a partial analytical form of the CD driving for the lattice with two excitations. 
Our numerical results demonstrate that circuit errors and environmental noise have negligible effects on our scheme under practical parameters. We also show that our scheme can be characterized through the detection of qubit operators. 
This approach can lead to a promising pathway to high-fidelity state preparation in a significantly reduced timescale when compared to conventional adiabatic methods. 
\end{abstract}
\maketitle

\section{Introduction}
Adiabatic evolution involves a gradual transformation of a quantum system from an initial state to a desired final state by slowly varying a time-dependent Hamiltonian, and has been extensively studied for quantum state preparation in quantum information processing~\cite{Albash2018,FarhiScience2001}. 
In noisy quantum systems~\cite{Preskill_2018, Bharti_2021}, achieving high fidelity for the prepared states using the adiabatic approach requires a balance in choosing the evolution time: it should be long enough to reduce unwanted diabatic transitions yet short enough to mitigate decoherence from environmental noise.  
Various approaches have been used to develop quantum shortcut to adiabaticity, where reverse engineering of diabatic transitions is employed to accelerate a slow adiabatic process via nonadiabatic shorcuts~\cite{Guery-Odelin2019, Kolodrubetz2017, UnanyanOptCommon1997,  EmmanouilidouPRL2000}. These approaches include counter-diabatic (CD) driving~\cite{Berry2009,Demirplak2008, Chen:PhysRevLett:2010:105, delCampoPRL2012, Damski2014}, invariant-based approaches~\cite{Chen:PhysRevLett:2010:104,Jing2013}, the derivative removal of adiabatic gate (DRAG) approach~\cite{Motzoi2009}, dressed state approaches~\cite{Baksic2016} \emph{etc.}. Quantum shortcut to adiabaticity has been proposed and experimentally demonstrated in many quantum systems, such as superconducting qubits, defect qubits, quantum dot arrays, and cold atoms~\cite{Wang2018, Vepsalainen2018, Theis2016, Martinis2014, Yan2019, Liang2016, Du2016, Zhou2017, Masuda2018, Bason:NatPhys:2011, Hatomura2018, Stefanatos2018,Zhang2015}. 
One hurdle that prevents the general application of these quantum shortcut approaches is the complexity of the Hamiltonian required for reverse engineering, which often involves multipartite or nonlocal interactions. A number of approaches have been explored to overcome this hurdle by deriving local CD driving, including approach that explores the self-similar nature of a quantum system, mean-field approximation, variational approach, Floquet engineering, digitalized approach, and optimal control approach~\cite{delCampoPRL2013, DeffnerPRX2014, Sels2017,Takahashi2017, Hegade2022L,Opatrny2014, BoyersPRA2019, ClaeysPRL2019, KeeverPRXQuantum2024, CepaitePRXQuantum2023}. 

Jaynes-Cummings (JC) lattices~\cite{Hartmann:2006, Greentree:2006, Angelakis:2007, 2007RossiniPRL_JC, 2008NeilPra_BH, 2009KochPra_QS, Seo2015:1, Xue2017, Noh2017Review} have been studied in various experimental systems, including superconducting qubits coupled to cavities, solid-state defects coupled to nanocavities, and trapped ions~\cite{2012HouckNP_JCQS, Hoffman:2011, Sala2015Nanophotonics, Lepert2011Atom, Ivanov2009Ion, Toyoda2013Ion, Debnath2018Ion, BWLi2022Ion}, and can exhibit quantum or dissipative phase transitions between the Mott-insulating and superfluid phases in the thermodynamic limit~\cite{HouckPRX2017, KeelingPRL2012,Seo2015:1, Xue2017}. The states of finite-sized JC lattices can also carry related features, resulting in novel phenomena such as the photon blockade effect and novel entanglement when properly populated with polariton excitations~\cite{Larson2022Review, TianPRL2011}. A key step in demonstrating these effects is to prepare desired quantum states with a finite number of excitations. In previous works, we employed an adiabatic approach and a quantum optimal control approach for state preparation in this system~\cite{KCaiNpj2021, ParajuliSciRep2023}. 

In our current work, we present a quantum shortcut to adiabaticity for state preparation in a finite-sized JC lattice by applying CD driving during an adiabatic evolution. In our approach, we explore the symmetry of the system's eigenstates to convert the CD driving to an implementable Hamiltonian for two-site and three-site lattices. We find that for a two-site lattice with one excitation, the CD driving can involve only local qubit-cavity couplings, and with two excitations, the CD driving can have a partial analytical form that involves four-operator couplings in the antisymmetric subspace. We also study the effects of control errors and environmental noise on our scheme, which complements previous CD driving studies on noisy and open quantum systems~\cite{BoyersPRA2019, FunoNoriPRL2021, AlipourQuantum2020}.
Our numerical results demonstrate that the effects of circuit errors and environmental noise are negligible under practical parameters.
Meanwhile, this scheme can be characterized by performing measurement on the qubits.
This work hence presents an implementable CD driving that can greatly shorten the evolution time of an adiabatic process in a JC lattice and provides a promising avenue to achieving high-fidelity state preparation in a significantly reduced timescale. 

\section{Jaynes-Cummings Lattice}
\begin{figure}[t]
\includegraphics[clip, width=8.5cm]{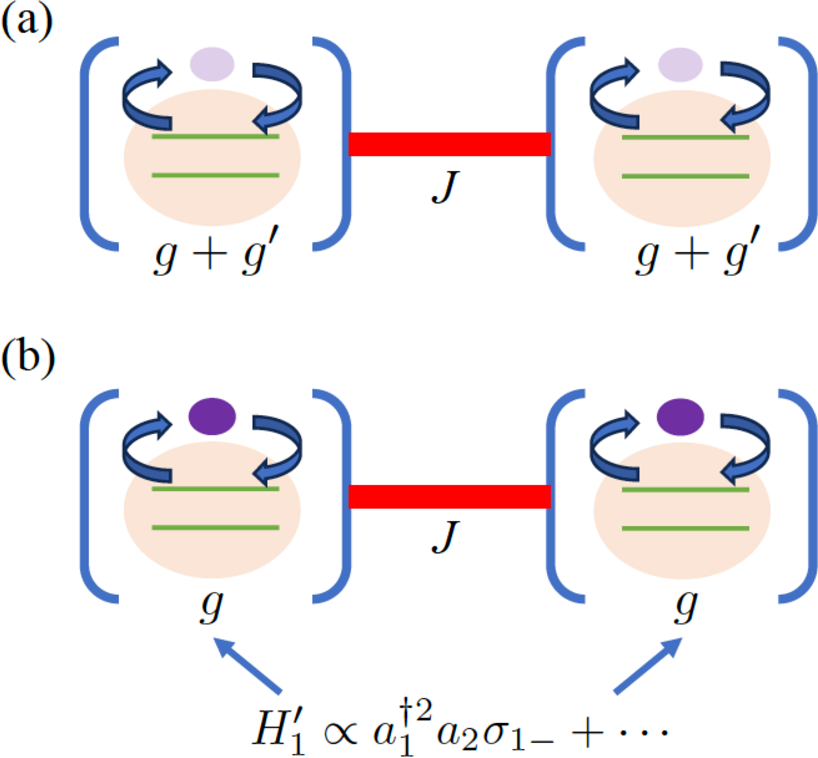}
\caption{Schematic of a two-site JC lattice with (a) one polariton excitation and (b) two polariton excitations. Each site is composed of a qubit (two-level system) coupled to a cavity mode with coupling strength $g$, and the cavities are connected by photon hopping with rate $J$. In (a), the CD driving includes only local qubit-cavity couplings denoted by $g^\prime$; in (b), the CD driving is shown as $H_{1}^\prime$.}
\label{fig1}
\end{figure}

Consider a finite-sized JC lattice composed of JC models coupled via photon hopping between adjacent sites, as illustrated in Fig.~\ref{fig1}. The Hamiltonian of the JC lattice can be written as  $H_{\rm JC} = H_{0} + g V_{g} + J V_{J}$, where $H_{0}=\sum_{j} ( \omega_0 a_{j}^{\dagger}a_{j} + \frac{\omega_z}{2} \sigma_{jz})$ is the Hamiltonian of the uncoupled qubits and cavity modes, $V_g= \sum_{j} ( a_{j}^{\dagger}\sigma_{j-}+\sigma_{j+}a_j)$ describes the onsite qubit-cavity coupling, and $V_J = - \sum_{j} (a_{i}^{\dagger}a_{i+1}+a_{i+1}^{\dagger}a_{i})$ describes the photon hopping between neighboring sites. Here, $a_j$ ($a_j^\dagger$) is the annihilation (creation) operator of the cavity modes, $\sigma_{j\pm}$, $\sigma_{jz}$ are the Pauli operators of the qubits, $\omega_0$ is the frequency of the cavities, $\omega_z$ is the qubit energy splitting, $g$ is the strength of the qubit-cavity coupling, $J$ is the photon hopping rate,  and  $j\in [1,\, N]$ with $N$ being the number of sites. We have assumed $\hbar=1$ for convenience of discussion. 
In the rotating frame defined by $H_{0}^{\rm rot}=\omega_{z}\sum_{j}(a_{j}^{\dagger}a_{j} + \frac{1}{2}\sigma_{jz})$, the Hamiltonian of the JC lattice becomes $H_{\rm r} = \sum_j \Delta  a_{j}^\dagger a_{j}   + g V_{g} + J V_{J}$, where $\Delta=\omega_0-\omega_z$ is the detuning between the cavity modes and the qubits.

We denote the basis states of a single JC model as $\vert n,s\rangle$, with $n\ge0$ being the photon number of the cavity mode and $s=g,\,e$ being the ground and excited states of the qubit, respectively. The eigenstates of a single JC model include the ground state $\vert 0,g\rangle$ and the doublets $\vert n,\pm\rangle$ ($n\ge1$), which are superpositions of the basis states $\vert n,g\rangle $ and $\vert n-1, e\rangle$ and correspond to eigenstates with $n$ polariton excitations. 
In the thermodynamic limit with $N\rightarrow \infty$, a JC lattice can exhibit various many-body effects such as the quantum and dissipative phase transitions between the Mott insulating and the superfluid phases~\cite{Hartmann:2006, Greentree:2006, Angelakis:2007}. It can be shown that the ground states of a finite-size JC lattice exhibit similar features to these phases~\cite{Seo2015:1, KCaiNpj2021, ParajuliSciRep2023}. 

A key step in studying novel phenomena in a JC lattice is the preparation of desired quantum states for a finite number of polariton excitations. One approach to generate such states is adiabatic evolution~\cite{ KCaiNpj2021}, where the system is first prepared in the ground state of an initial Hamiltonian $H_{\rm r}(t_0)$ at time $t_0$, followed by adiabatic tuning of the system parameters to reach the target Hamiltonian $H_{\rm r}(T)$ at the final time $T$. The ground state of the target Hamiltonian is designed to be the desired state, which is often hard to prepare directly by gate operations. In the end of the adiabatic evolution, the system will reach the desired state. 
However, the adiabaticity of the evolution and hence the fidelity of the prepared states can be impaired by unwanted diabatic transitions to excited states. Below, we will employ a quantum shortcut to adiabaticity approach to study the robust generation of quantum states in a JC lattice within a significantly reduced timeframe. 

\section{Counter-Diabatic Driving\label{sec:CD}}
CD driving can be applied to the Hamiltonian $H_{\rm r}(t)$ to cancel diabatic transitions and generate quantum shortcut to adiabaticity during an adiabatic evolution. The CD driving has the general form~\cite{Berry2009, Demirplak2008,Guery-Odelin2019}: 
\begin{equation}
H_{1}(t)=i\sum_{m\neq n}\frac{|m(t)\rangle\langle m(t)|\partial_{t}H_{\rm r}|n(t)\rangle\langle n(t)|}{E_{n}(t)-E_{m}(t)},\label{eq:H1}
\end{equation}
where $\vert m\rangle$, $\vert n\rangle$ are instantaneous eigenstates of $H_{\rm r}(t)$ at time $t$ and $E_{m}$, $E_{n}$ are the corresponding eigenenergies. The total Hamiltonian of this system then becomes $H_{\rm tot}=H_{\rm r}(t)+H_{1}(t)$. The exact CD Hamiltonian in (\ref{eq:H1}) often includes nonlocal or multipartite interactions due to the complexity of the quantum system of interest. Hence, it is challenging to implement this Hamiltonian in systems with practical parameters. 

In this section, we convert the CD driving in (\ref{eq:H1}) to an implementable Hamiltonian for adiabatic evolutions in two-site and three-site JC lattices. Our results show that the CD driving can comprise only local qubit-cavity couplings and is implementable using current technology. 

\subsection{Two sites one excitation \label{ssec:S2E1}}
\emph{Eigenstates.} We first study a two-site JC lattice ($N=2$) with only one polariton excitation. The allowable Hilbert space for this system includes four basis states: $|1,g\rangle_1\vert 0,g\rangle_2$, $|0,e\rangle_1\vert0,g\rangle_2$, $|0,g\rangle_1\vert 1,g\rangle_2$, and $|0,g\rangle_1\vert0,e\rangle_2$, where the subscripts refer to the sites 1 and 2 in the lattice. Written in terms of this basis set, the Hamiltonian has the form: 
\begin{equation}
H_{\rm r}=\begin{pmatrix}
\Delta & g & -J & 0\\
g & 0 & 0 & 0\\
-J & 0 & \Delta & g\\
0 & 0 & g & 0
\end{pmatrix}
\label{eq:Hr}
\end{equation}
for given values of $g,\,J,$ and $\Delta$. The eigenvalues $E_{n}$ ($n\in [1,4]$) of this system are 
\begin{equation}
    E_{1,2} =\frac{1}{2}(\Delta_{J}^{-}\mp\chi_{1}^{-}), \quad E_{3,4} =\frac{1}{2}(\Delta_{J}^{+}\mp\chi_{1}^{+}) \label{eq:e1}
\end{equation}
with $\Delta_J^{\pm}=\Delta\pm J$ and $\chi_{1}^{\pm}=\sqrt{(\Delta_{J}^{\pm})^{2}+4g^{2}}$. 
The corresponding eigenvectors are
\begin{equation}
\begin{aligned}
v_{1}&=\frac{1}{2}\begin{pmatrix}
-\sqrt{\frac{\chi_{1}^{-}-\Delta_{J}^{-}}{\chi_{1}^{-}}}\\
\sqrt{\frac{\chi_{1}^{-}+\Delta_{J}^{-}}{\chi_{1}^{-}}}\\
-\sqrt{\frac{\chi_{1}^{-}-\Delta_{J}^{-}}{\chi_{1}^{-}}}\\
\sqrt{\frac{\chi_{1}^{-}+\Delta_{J}^{-}}{\chi_{1}^{-}}}
\end{pmatrix}, \quad
v_{2}=\frac{1}{2}\begin{pmatrix}
\sqrt{\frac{\chi_{1}^{-}+\Delta_{J}^{-}}{\chi_{1}^{-}}}\\
\sqrt{\frac{\chi_{1}^{-}-\Delta_{J}^{-}}{\chi_{1}^{-}}}\\
\sqrt{\frac{\chi_{1}^{-}+\Delta_{J}^{-}}{\chi_{1}^{-}}}\\
\sqrt{\frac{\chi_{1}^{-}-\Delta_{J}^{-}}{\chi_{1}^{-}}}
\end{pmatrix}, \\
 v_{3}&=\frac{1}{2}\begin{pmatrix}
\sqrt{\frac{\chi_{1}^{+}-\Delta_{J}^{+}}{\chi_{1}^{+}}}\\
-\sqrt{\frac{\chi_{1}^{+}+\Delta_{J}^{+}}{\chi_{1}^{+}}}\\
-\sqrt{\frac{\chi_{1}^{+}-\Delta_{J}^{+}}{\chi_{1}^{+}}}\\
\sqrt{\frac{\chi_{1}^{+}+\Delta_{J}^{+}}{\chi_{1}^{+}}}
\end{pmatrix}, \quad
 v_{4}=\frac{1}{2}\begin{pmatrix}
-\sqrt{\frac{\chi_{1}^{+}+\Delta_{J}^{+}}{\chi_{1}^{+}}}\\
-\sqrt{\frac{\chi_{1}^{+}-\Delta_{J}^{+}}{\chi_{1}^{+}}}\\
\sqrt{\frac{\chi_{1}^{+}+\Delta_{J}^{+}}{\chi_{1}^{+}}}\\
\sqrt{\frac{\chi_{1}^{+}-\Delta_{J}^{+}}{\chi_{1}^{+}}}
\end{pmatrix}.
\end{aligned} 
\label{eq:v1}
\end{equation}
The eigenstate $v_1$ is the ground state with energy $E_1$.
From (\ref{eq:v1}), we observe that the states $v_1$ and $v_2$ are symmetric with respect to the exchange of the states on the two sites, while the states $v_3$ and $v_4$ are antisymmetric and acquire a factor of $-1$ when the states on the two sites are swapped.  
\begin{figure}[t]
\begin{center}
\includegraphics[clip, width=8.5cm]{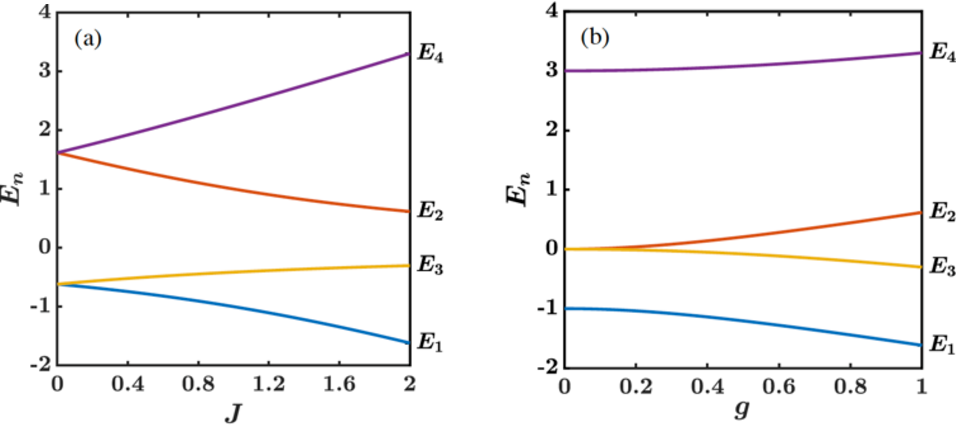} 
\caption{Eigenenergies $E_n$ ($n\in[1,4]$) of a two-site JC lattice with one excitation. (a) $E_n$ vs hopping rate $J$ for $g=1$ and $\Delta=1$. (b) $E_n$ vs coupling constant $g$ for $J=2$ and $\Delta=1$. All parameters are in dimensionless units.}
\label{fig2}
\end{center}
\end{figure}
In Fig.~\ref{fig2}(a), we plot the energy spectrum of the system as a function of the photon hopping rate $J$ for $g=1$ and $\Delta=1$. At $J=0$, $\Delta_J^{\pm}=\Delta$, resulting in $E_1=E_3$ ($E_2=E_4$), which indicates that the states $v_1$ and $v_3$ ($v_2$ and $v_4$) are degenerate.
In Fig.~\ref{fig2}(b), we plot the energy spectrum vs the qubit-cavity coupling $g$ for $J=2$ and $\Delta=1$. At $g=0$ and $\Delta<J$, $E_2=E_3=0$, and $v_2$ and $v_3$ are degenerate.

 We want to emphasize that the symmetry of these eigenstates is crucial for our development of an implementable CD driving that only involves local qubit-cavity couplings. During an adiabatic evolution, the variation of the Hamiltonian does not induce transition of the system from a symmetric to an antisymmetric state due to the inherent symmetry of the Hamiltonian.
Furthermore, this symmetry also facilitates the preparation of the ground state $v_1$ at $J=0$ via resonant pumping, without inducing excitation to the state $v_3$ that is degenerate to $v_{1}$.

\emph{Ramping of hopping rate $J$.}  Assuming that the hopping rate $J(t)$ is continuously tuned from $J(0)=0$ at time $t=0$ to $J(T)=J_f$ at the final time $T$ with $J_f$ being the target hopping rate, while keeping the qubit-cavity coupling and the detuning unchanged during the adiabatic process. At $t=0$, the system is prepared in the ground state $v_1$ for $J=0$ by resonantly pumping the system from the state $\vert 0,g\rangle_1 \vert 0,g\rangle_2$. The variation of $J(t)$ induces off-diagonal matrix elements between the instantaneous eigenstates, resulting in undesired transitions to the excited states.
The nonzero matrix elements are: $\vert \langle v_{1}|V_J|v_{2}\rangle\vert =g/ \chi_{1}^{-}$, $\vert \langle v_{3}|V_J|v_{4}\rangle\vert = g/ \chi_{1}^{+}$, along with their conjugate elements, which induce transitions between the states $v_1$ and $v_2$, and between the states $v_3$ and $v_4$, respectively. We then derive the CD driving using (\ref{eq:H1}): 
\begin{eqnarray}
H_{1}(t)&=&i\frac{g}{(\chi_{1}^{-})^{2}}\frac{dJ}{dt}(|v_{1}\rangle\langle v_{2}|-|v_{2}\rangle\langle v_{1}|) \nonumber\\
&&-i\frac{g}{(\chi_{1}^{+})^{2}}\frac{dJ}{dt}(|v_{3}\rangle\langle v_{4}|-|v_{4}\rangle\langle v_{3}|). 
\label{eq:H11}
\end{eqnarray}
Hence, when the system is in the ground state $v_1$, the only allowable transition is to the symmetric eigenstate $v_2$. As discussed above, this is due to the symmetry of the eigenstates and the Hamiltonian. A variation of the hopping rate in the Hamiltonian preserves the exchange symmetry of the state, which can only induce diabatic transition to the state $v_2$. Consequently, the state at an arbitrary time $t$ during the evolution remains symmetric and can only be a superposition of the instantaneous $v_{1}$ and $v_{2}$ states. Thus, the second term in $H_{1}(t)$ does not actively contribute to the elimination of diabatic transitions, and its amplitude can be adjusted with flexibility, without affecting the adiabatic dynamics of this system. Below, we will construct an easier-to-implement CD driving using this property.

We first convert the CD driving in (\ref{eq:H11}) to physical operators of the qubits and cavities to see if it can be easily implemented. In the Hilbert space for one polariton excitation, we have 
$a_{1}^{\dagger}\sigma_{1-}=|1,g\rangle_1\langle 0,e\vert \otimes \vert0,g\rangle_2\langle 0,g|$, $a_{1}^{\dagger}a_{2}=|1,g\rangle_1 \langle 0,g\vert\otimes \vert 0,g\rangle_2\langle 1,g|$, and so forth. 
Using the expression of the eigenstates in (\ref{eq:v1}), we derive:
\begin{subequations}
\begin{align}
|v_{1}\rangle\langle v_{2}|-|v_{2}\rangle\langle v_{1}|=&\frac{1}{2}\left( S_{+}^\dagger A_{+} -A_{+}^\dagger S_{+} \right), 
\label{eq:v1v2}\\
|v_{3}\rangle\langle v_{4}|-|v_{4}\rangle\langle v_{3}|=&\frac{1}{2}\left(S_{-}^\dagger A_{-} -A_{-}^\dagger S_{-}\right),
\label{eq:v3v4}
\end{align}
\end{subequations}
with $A_{\pm} = a_1 \pm a_2 $ and $S_{\pm} = \sigma_{1-}\pm\sigma_{2-}$.  And the CD driving is
\begin{eqnarray}
H_{1}(t) &=&i\frac{g}{2(\chi_{1}^{-})^{2}}\frac{dJ}{dt}\left( S_{+}^\dagger A_{+} -A_{+}^\dagger S_{+} \right) \nonumber \\
&& -i\frac{g}{2(\chi_{1}^{+})^{2}}\frac{dJ}{dt} \left(S_{-}^\dagger A_{-} -A_{-}^\dagger S_{-}\right). \label{eq:H1operator_J}
\end{eqnarray}
This Hamiltonian comprises local couplings between qubits and cavities on the same site such as $\sigma_{1+}a_1$ and $\sigma_{2+}a_2$ as well as nonlocal couplings between qubits and neighboring cavities such as $\sigma_{1+}a_2$ and $\sigma_{2+}a_1$. 

By changing the coefficient of the second term in (\ref{eq:H1operator_J}) from $-i g/2(\chi_{1}^{+})^{2}$ to $i g/2(\chi_{1}^{-})^{2}$, we obtain the following CD Hamiltonian:  
\begin{equation}
    H_{1}^{\prime}(t)=ig^\prime\left(a_{1}\sigma_{1+}-a_{1}^{\dagger}\sigma_{1-}+a_{2}\sigma_{2+}-a_{2}^{\dagger}\sigma_{2-}\right).\label{eq:H1p_simp}
\end{equation}
This Hamiltonian includes only local qubit-cavity couplings with coupling strength given by $g^\prime=\frac{g}{(\chi_{1}^{-})^{2}}\frac{dJ}{dt}$. It can be interpreted as a Jaynes-Cummings coupling with a $e^{i\pi/2}$ phase, and can be implemented by adjusting local qubit-cavity couplings.

\begin{figure}[t]
\begin{center}
\includegraphics[clip,width=8.5cm]{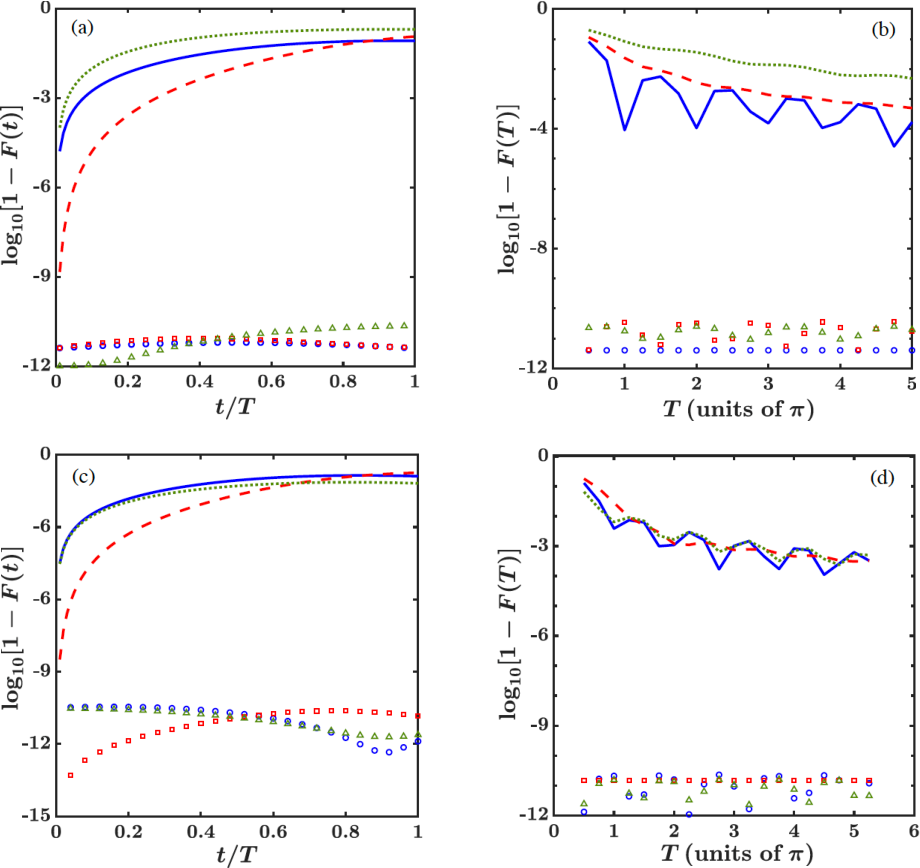}
\caption{(Color online) (a, c) Infidelity $1-F(t)$ vs $t/T$ for a total evolution time of $T=0.5 \pi$ for two-site and three-site JC lattices, respectively. (b, d) Infidelity $1-F(T)$ vs $T$ for two-site and three-site JC lattices, respectively. 
In all plots, blue solid line (blue circles): linear ramping of hopping rate $J$ for $g\equiv 1$, $\Delta\equiv 1$, $J(0)=0$ and $J_f=2$ under the adiabatic Hamiltonian $H_{\rm r}$ only (with the CD driving $H_1^\prime$ applied); red dashed line (red squares): quadratic ramping of $J$ for $g\equiv 1$, $\Delta\equiv 1$, $J(0)=0$ and $J_f=2$ under $H_{\rm r}$ (with $H_1^\prime$); and green dotted line (green triangles): linear ramping of coupling $g$ for $J\equiv 2$, $\Delta\equiv 1$, $g(0)=0$ and $g_f=1$ under $H_{\rm r}$ (with $H_1^\prime$). All parameters are in dimensionless units. }
\label{fig3}
\end{center}
\end{figure}

For demonstration, we simulate the time evolution of this system under the adiabatic Hamiltonian $H_{\rm r}$ with either $H_1$ or $H_1^\prime$ applied. We consider both linear ramping of the hopping rate with $J(t)=J_f \frac{t}{T}$ and quadratic ramping of the hopping rate with $J(t)=J_f\left(\frac{t}{T}\right)^2$. For linear ramping, $dJ/dt= J_f /T$; and for quadratic ramping, $dJ/dt= 2J_f t/T^2$. We choose the target hopping rate $J_f=2$, the qubit-cavity coupling $g\equiv1$, the detuning $\Delta\equiv1$, and the total time $T=0.5 \pi$.
We define the fidelity of the quantum state at time $t$ as $F(t)=|\langle v_1 (t)|v(t)\rangle|^{2}$, where $v(t)$ is the state at time $t$ and $v_1(t)$ is the instantaneous ground state of the adiabatic Hamiltonian at time $t$.
In Fig.~\ref{fig3}(a), we plot the infidelity $1-F(t)$ vs $t/T$. Without the CD driving, the infidelity increases significantly during the evolution, as the diabatic transitions can be serious for a very short evolution time of $T=0.5\pi$. In contrast, when the CD driving $H_{1}^{\prime}$ is applied, the infidelity, up to a small numerical error below $10^{-10}$, is zero throughout the evolution, and the system is preserved in the ground state. 
In Fig.~\ref{fig3}(b), we plot the infidelity $1-F(T)$ of the final state vs the total evolution time $T$. Without the CD driving, the infidelity can be quite high for a short evolution time. This result demonstrates that the CD approach can ensure high fidelity for a short evolution time, which is crucial for devices in the noisy intermediate-scale quantum (NISQ) regime~\cite{Preskill_2018}. 
As shown in Fig.~\ref{fig7}(a, b) in Appendix A, the infidelities under $H_1$ and $H_1^\prime$ are the same up to a small numerical error below $10^{-10}$, which confirms our analysis that these two CD drivings yield the same dynamics for the initial state $v_{1}$.

\emph{Ramping of qubit-cavity coupling $g$.} Now consider an adiabatic trajectory where the qubit-cavity coupling is ramped linearly with $g(t)=g_f t/T$ and $g_f$ being the target coupling strength. The nonzero transition matrix elements caused by the variation of $g$ are $\vert \langle v_{1}|V_g|v_{2}\rangle\vert = \Delta_{J}^{-}/\chi_{1}^{-}$, $\vert \langle v_{3}|V_g|v_{4}\rangle\vert =\Delta_{J}^{+}/\chi_{1}^{+} $, and their conjugate elements, which induce transitions between the states $v_1$ and $v_2$ as well as between the states $v_3$ and $v_4$, respectively, due to the symmetry of the eigenstates.
The CD driving can be derived as:
\begin{eqnarray}
H_{1}(t)&=&i\frac{\Delta_{J}^{-}}{2(\chi_{1}^{-})^{2}}\frac{dg}{dt}\left( S_{+}^\dagger A_{+} -A_{+}^\dagger S_{+} \right) \nonumber \\
&&-i\frac{\Delta_{J}^{+}}{2(\chi_{1}^{+})^{2}}\frac{dg}{dt}\left(S_{-}^\dagger A_{-} -A_{-}^\dagger S_{-}\right).\label{eq:H1operator_g}
\end{eqnarray}
Similar to (\ref{eq:H1operator_J}), this Hamiltonian includes nonlocal couplings between qubits and neighboring cavities. 
As discussed above, when  $\vert v_1\rangle$ is the initial state, only the state $\vert v_2\rangle$ could be excited by diabatic transitions. Therefore, we can vary the coefficient of the second term in (\ref{eq:H1operator_g}). By changing this coefficient from $-i\Delta_{J}^{+}/2(\chi_{1}^{+})^{2}$ to $i\Delta_{J}^{-}/2(\chi_{1}^{-})^{2}$, we obtain the CD driving:
\begin{equation}
H_{1}^\prime(t)=i g^\prime\left(a_{1}\sigma_{1}^{+}-a_{1}^{\dagger}\sigma_{1}^{-}+a_{2}\sigma_{2}^{+}-a_{2}^{\dagger}\sigma_{2}^{-}\right),
\label{eq:H1p_g_simple}
\end{equation}
which only involves local qubit-cavity couplings with strength $g^\prime = \frac{\Delta_{J}^{-}}{(\chi_{1}^{-})^{2}}\frac{dg}{dt}$. Consequently, implementing both the adiabatic Hamiltonian and the CD driving only requires the tuning of the local qubit-cavity coupling, which can significantly simplify the experimental setup.

We perform a numerical simulation on this system with hopping rate of $J\equiv2$, detuning of $\Delta\equiv1$, and a target coupling strength of $g_f=1$. At $t=0$ and $g(0)=0$, the initial state is $(|1,g\rangle_1\vert 0,g\rangle_2+|0,g\rangle_1\vert 1,g\rangle_2)/\sqrt{2}$, where the excitation is stored in the cavity modes.
The infidelity $1-F(t)$ vs $t/T$ for $T=0.5\pi$ is plotted in Fig.\ref{fig3}(a), and the infidelity at the final time $1-F(T)$ vs the total time $T$ is plotted in Fig.\ref{fig3}(b). The results are similar to the case of ramping the hopping rate $J$.

\emph{Self-protected trajectory.} The magnitude of the CD driving in (\ref{eq:H1p_g_simple}) is proportional to $\Delta_{J}^{-}=\Delta-J$. In the special case of $\Delta=J$, $\Delta_{J}^{-}=0$, leading to the interesting result: $H_1^\prime=0$, which means that no CD driving is required. It can be shown that the matrix element $\vert \langle v_{1}|V_g|v_{2}\rangle\vert \equiv 0$ under this condition, hence the only allowable diabatic transition for the initial state $v_{1}$ disappears. The Hamiltonian $H_{\rm r}$ is thus self-protected from diabatic transition, i.e., there is no diabatic transition when ramping the qubit-cavity coupling $g$. The ground state is $\frac{1}{\sqrt{2}}\left( \vert 1,-\rangle_1\vert 0,g\rangle_2 +\vert 0,g\rangle_1 \vert 1,-\rangle_2\right)$ throughout the evolution, with the system occupying the lower polariton state $\vert 1,-\rangle$ in one of the lattice sites.  

\subsection{Three sites one excitation\label{ssec:S3E1}}
We now study a three-site lattice with one polariton excitation under the open boundary condition with the photon hopping term $V_{J}= -(a_{1}^{\dagger}a_{2} + a_{2}^{\dagger}a_{3} + \text{h.c.})$, where sites $1$ and $3$ possess mirror reflection symmetry.  
The allowable Hilbert space comprises six basis states: $|1,g\rangle_1 \vert 0,g\rangle_2 \vert 0,g\rangle_3$, $|0,e\rangle_1 \vert 0,g\rangle_2 \vert 0,g\rangle_3$, $|0,g\rangle_1 \vert 1,g\rangle_2 \vert 0,g\rangle_3$,  $|0,g\rangle_1 \vert 0,e \rangle_2 \vert 0,g\rangle_3$,  $|0,g\rangle_1 \vert0,g\rangle_2 \vert 1,g\rangle_3$, and $|0,g\rangle_1 \vert 0,g \rangle_2 \vert 0,e\rangle_3$. 
The Hamiltonian in terms of this basis set has the form:
\begin{equation}
H_{\rm r} =  \begin{pmatrix}
\Delta  & g & -J & 0 & 0 & 0\\
g & 0 & 0 & 0 & 0 & 0\\
-J & 0 & \Delta & g & -J & 0\\
0 & 0 & g & 0 & 0 & 0\\
0 & 0 & -J & 0 & \Delta & g\\
0 & 0 & 0 & 0 & g & 0
\end{pmatrix}.
\end{equation}
We numerically calculate the eigenstates of this system. Four eigenstates exhibit symmetry regarding the exchange of the states on sites $1$ and $3$. For $g=1$, $J=2$, and $\Delta=1$, these symmetric states include the ground state $v_1$, the second excited state $v_3$, the third excited state $v_4$, and the highest state $v_6$. When ramping the hopping rate $J$, diabatic transitions occur only between $v_1$ and $v_4$ and between $v_3$ and $v_6$. Thus, the CD driving takes the form of $H_1 =\alpha_1 \vert v_4\rangle \langle v_1 \vert +\alpha_2 \vert v_6 \rangle \langle v_3 \vert + \text{h.c.}$ with coefficients $\alpha_{1}$ and $\alpha_{2}$. When expressed in terms of the physical operators, we have:
\begin{eqnarray}
    H_{1} &=& i g_{n} \left[\left(a_1^\dagger+a_3^\dagger\right)\sigma_{2-} + a_2^\dagger\left(\sigma_{1-}+\sigma_{3-}\right) - h.c. \right] \nonumber \\
    && + i g_{l}  \left(a_1^\dagger \sigma_{3-} + a_3^\dagger\sigma_{1-} - h.c.\right)  \nonumber \\
    && + i g_{l} \left( a_1^\dagger\sigma_{1-} + 2 a_2^\dagger\sigma_{2-} +a_3^\dagger\sigma_{3-} - h.c. \right)    \label{eq:H1S3E1}
\end{eqnarray}
which includes nonlocal couplings with strengths $g_{n}$ and $g_{l}$, such as the coupling between the cavities on sites $1$ and $3$ and the qubit on site $2$, as well as local couplings with strength $g_{l}$ for sites $1,\,3$ and strength $2g_{l}$ for site $2$. 

We employ the same approach as in Sec.~\ref{ssec:S2E1} to convert the CD Hamiltonian to a form that is easier to implement. As discussed above, for initial state $v_{1}$, ramping of the hopping terms only induces transition to the state $v_{4}$. Therefore, the coefficient $\alpha_2$ in $H_{1}$ can be varied with flexibility without affecting the dynamics of this system. By adjusting $\alpha_2$, we find 
\begin{eqnarray}
    H_{1}^\prime &=& i g^\prime \left( a_1^\dagger\sigma_{1-} + 2 a_2^\dagger\sigma_{2-} +a_3^\dagger\sigma_{3-} - h.c. \right)
    \nonumber \\
   && + i g^\prime \left(a_1^\dagger \sigma_{3-} + a_3^\dagger\sigma_{1-} - h.c.\right),  \label{eq:H1S3E1_simp}
\end{eqnarray}
where the couplings between the qubit (cavity) on site $2$ and the cavities (qubits) on sites $1$ and $3$ are no longer required. This Hamiltonian still includes nonlocal terms between sites $1$ and $3$ with coupling strength $g^\prime$. This Hamiltonian can be implemented for a small lattice of three sites by coupling the qubits and cavities on sites $1$ and $3$. However, this result shows that we cannot always find a local form of the CD driving. In Fig.~\ref{fig3}(c, d), we plot the infidelity of the prepared state under the adiabatic Hamiltonian $H_{\rm r}$ both without and with the CD driving $H_{1}^\prime$. The result is similar to that of the two-site lattice discussed in Sec.~\ref{ssec:S2E1}.

\subsection{Two sites two excitations \label{ssec:S2E2}}
\begin{figure}[t]
\includegraphics[clip,width=8.5cm]{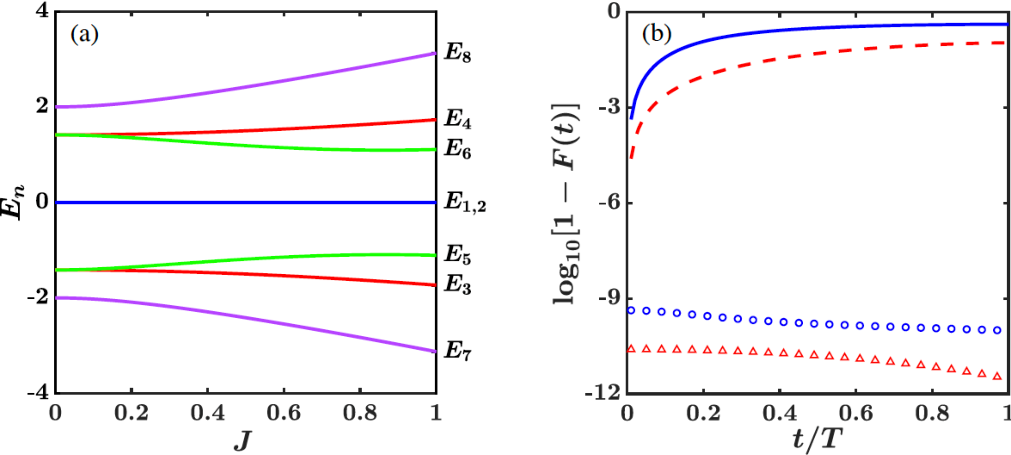} 
\caption{(Color online) (a) Eigenenergies $E_n$ ($n\in[1,8]$) vs hopping rate $J$. (b) Infidelity $1-F(t)$ vs $t/T$ for a total evolution time of $T=0.5 \pi$. 
Blue solid line (blue circles): linear ramping of hopping rate $J$ under the adiabatic Hamiltonian $H_{\rm r}$ only (with the CD driving $H_1$ applied) with the initial state being the ground state $v_{7}$ at $J=0$; and red dashed line (red squares): linear ramping of $J$ under $H_{\rm r}$ (with $H_1^{\prime}$) with the initial state being the first excited state $v_{3}$ at $J=0$. 
Here, $g\equiv 1$, $\Delta\equiv 0$, $J(0)=0$, $J_f=1$, and all parameters are in dimensionless units.}
\label{fig4}
\end{figure}

\emph{Eigenstates.} For a two-site JC lattice with two polariton excitations, the Hilbert space includes eight basis states: $|2,g\rangle_1 \vert 0,g\rangle_2$, $|1,e\rangle_1 \vert 0,g\rangle_2$, $|0,g\rangle_1 \vert 2,g\rangle_2$, $|0,g\rangle_1 \vert 1,e\rangle_2$, $|1,g\rangle_1 \vert 1,g\rangle_2$, $|0,e\rangle_1 \vert 1,g\rangle_2$, $|1,g\rangle_1 \vert 0,e\rangle_2$, and $|0,e\rangle_1 \vert 0,e\rangle_2$. In the first four states, both excitations occupy one of the two sites, while in the last four states, one excitation is located on each site. Using this basis set, the Hamiltonian $H_{\rm r}$ can be expressed as:
\begin{equation}
H_{\rm r}=\begin{pmatrix}
2\Delta & \sqrt{2}g & 0 & 0 & -\sqrt{2}J & 0 & 0 & 0\\
\sqrt{2}g & \Delta & 0 & 0 & 0 & -J & 0 & 0\\
0 & 0 & 2\Delta & \sqrt{2}g & -\sqrt{2}J & 0 & 0 & 0\\
0 & 0 & \sqrt{2}g & \Delta & 0 & 0 & -J & 0\\
-\sqrt{2}J & 0 & -\sqrt{2}J  & 0 & 2\Delta & g & g &  0\\
0 & -J & 0 & 0 & g & \Delta & 0 & g\\
0 & 0 & 0 & -J & g & 0 & \Delta & g\\
0 & 0 & 0 & 0 & 0 & g & g & 0
\end{pmatrix}.
\label{eq:Htr2}
\end{equation}
For detuning $\Delta=0$, the eigenvalues of this Hamiltonian can be derived analytically as: 
\begin{subequations}
\begin{align}
E_{1,2}&=0,\\
E_{3,4}&=\mp \sqrt{2g^{2}+J^{2}},\\
E_{5,6}&=\mp \frac{\sqrt{6g^{2}+5J^{2}-\sqrt{4g^{4}+60g^{2}J^{2}+9J^{4}}}}{\sqrt{2}},\\
E_{7,8}&=\mp \frac{\sqrt{6g^{2}+5J^{2}+\sqrt{4g^{4}+60g^{2}J^{2}+9J^{4}}}}{\sqrt{2}}.
\end{align}
\label{eq:E1_8}
\end{subequations}
In Fig.~\ref{fig4}(a), we plot the eigenenergies vs the hopping rate $J$ at $g=1$ and $\Delta=0$. We denote the eigenstates as $v_i$ with $i\in[1,8]$, corresponding to the eigenvalues given in (\ref{eq:E1_8}). Here, $v_7$ is the ground state, while $v_{1}$ and $v_{2}$ are degenerate states with $E_{1,2}=0$.

\emph{Ramping of hopping rate $J$.} Assuming that the hopping rate is linearly ramped with $J=J_f t/T$, the target hopping rate $J_f=1$, $g\equiv 1$, and $\Delta\equiv0$. The CD driving $H_{1}$ can be obtained numerically using Eq.~(\ref{eq:H1}). We simulate the dynamics of this system with the initial state being the ground state $v_7$ at $J=0$ for a total evolution time of $T=0.5\pi$. In Fig.~\ref{fig4}(b), we plot the infidelity $1-F(t)$ of the quantum state at time $t$ as a function of $t/T$. The infidelity at the final time $T$ exceeds $0.4$ under the adiabatic Hamiltonian $H_{\rm r}$, whereas it remains negligible when the CD driving is applied. 

By examining the eigenstates of $H_{\rm r}$, we find that the eight eigenstates can be divided into two subsets. The first subset contains three states $\{v_{2},\, v_{3},\,v_{4}\}$; while the second subset contains the remaining five states, including the ground state $v_{7}$. When varying the hopping rate $J$, the only nonzero transition matrix elements are between states within the same subset, and there is no diabatic transition between states in different subsets. Thus, for an initial state in the first subset, we consider the CD driving $H_{1}^{\prime}$ that only eliminates the diabatic transitions between eigenstates in this subset. As detailed in Appendix B, we derive
\begin{equation}
    H_1^\prime(t) = i\frac{g}{2\left(J^{2}+2g^{2}\right)}\frac{d J}{d t}\left(
A_2^\dagger C_2 - C_2^\dagger A_2 \right), \label{eq:H1S2E2}
\end{equation}
where $A_2 = a_{1}^2 - a_{2}^2$ and $C_2= a_{2}\sigma_{1-}-a_{1}\sigma_{2-}$.
This Hamiltonian consists of four-operator terms, which can be formed through effective second-order couplings in the perturbative regime. 

To demonstrate the effectiveness of Eq.~(\ref{eq:H1S2E2}), we conducted a numerical simulation of the adiabatic process on the initial state $v_3$ from the first subset, which corresponds to the first excited state in Fig.~\ref{fig4}(a). As shown in Fig.~\ref{fig4}(b), up to a small numerical error below 
$10^{-10}$, the infidelity remains zero throughout the evolution, indicating that the diabatic transitions are completely cancelled by the CD driving $H_1^\prime$. In contrast, if the initial state is the ground state $v_{7}$ from the second subset, the infidelities both without and with the CD driving $H_{1}^{\prime}$ will be the same [see the blue solid line in Fig.~\ref{fig4}(b)]. These numerical findings clearly show that the CD driving $H_{1}^{\prime}$ is only effective for states in the first subset, but not for states in the second subset, which is consistent with our analysis above. 

\section{Error, Noise, and Measurement}
In this section, we discuss the effects of control errors and environmental noise on the adiabatic evolution under the CD driving, as well as the characterization of the CD approach through qubit measurements.

\begin{figure}[t]
\includegraphics[clip,width=8.5cm]{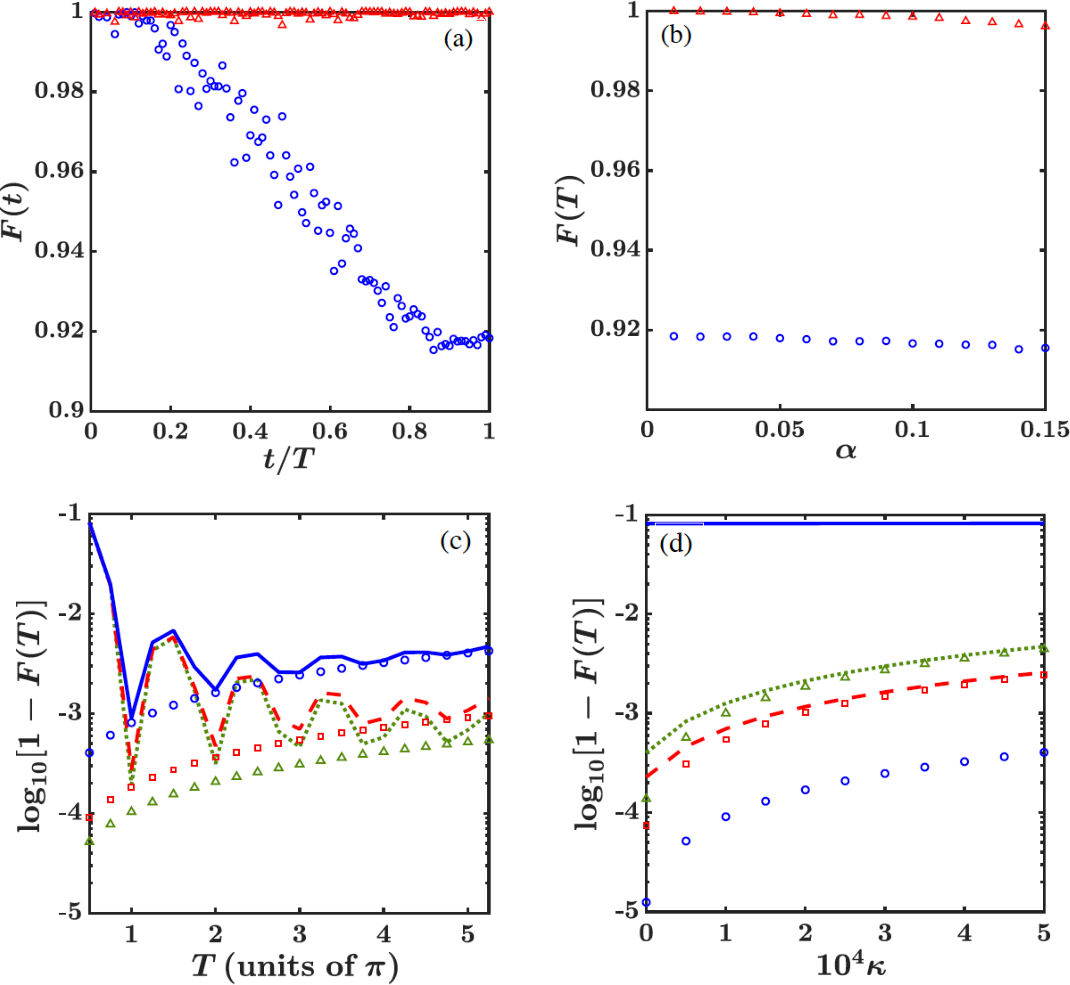}
\caption{(Color online) (a) Fidelity $F(t)$ vs $t/T$ for one sample of errors with $\alpha=0.05$. (b) Fidelity $F(T)$ vs $\alpha$ averaged over 100 samples of errors. In (a) and (b), blue circles (red triangles): under the adiabatic Hamiltonian $H_{\rm r}$ only (with the CD driving $H_1^\prime$ applied); and the total evolution time is $T=0.5\pi$. 
(c) Infidelity $1-F(T)$ vs $T$ for $\gamma=\frac{5}{\pi}\times 10^{-5}$. Blue solid line (blue circles): for $\kappa=5\times 10^{-4}$ under the adiabatic Hamiltonian $H_{\rm r}$ only (with the CD driving $H_1^\prime$ applied); red dashed line (red squares): for $\kappa=10^{-4}$ under $H_{\rm r}$ (with $H_1^\prime$); and green dotted line (green triangles): for $\kappa=5\times 10^{-5}$ under $H_{\rm r}$ (with $H_1^\prime$).
(d) Infidelity $1-F(T)$ vs $\kappa$ for $\gamma=\frac{5}{\pi}\times 10^{-5}$. Blue solid line (blue circles): for $T=0.5\pi$ under the adiabatic Hamiltonian $H_{\rm r}$ only (with the CD driving $H_1^\prime$ applied); red dashed line (red squares): for $T=3\pi$ under $H_{\rm r}$ (with $H_1^\prime$); and green dotted line (green triangles): for $T=5.5\pi$ under $H_{\rm r}$ (with $H_1^\prime$).
All plots are for linear ramping of hopping rate $J$ with parameters given in Fig.~\ref{fig3} in dimensionless units.}
\label{fig5}
\end{figure}

\emph{Control errors.} Adjusting the system parameters can introduce classical control errors, potentially impacting the performance of the CD approach. To investigate the effects of these errors, we conducted numerical simulations of the adiabatic process by introducing random fluctuations to the time-dependent coupling constants. 
Consider the linear ramping of the hopping rate $J(t)$ discussed in Sec.~\ref{ssec:S2E1}, where the CD driving only consists of time-dependent local qubit-cavity couplings. We introduce a fluctuation $\delta J(t)$ to the hopping rate with $J(t) \rightarrow J(t) + \delta J(t)$, and a fluctuation $\delta g(t)$ to the qubit-cavity coupling [including the CD terms in (\ref{eq:H1p_g_simple})] with $g(t) \rightarrow g(t) + \delta g(t)$. Here, $\delta J(t)$ and $\delta g(t)$ are Gaussian random numbers with a standard deviation $\alpha J_f$ and $\alpha g$, respectively, where $\alpha$ is the ratio of the fluctuation amplitude to the coupling strength. We choose $\alpha$ to be below $0.15$, i.e., the fluctuations are below $15\%$ of the coupling strengths. In superconducting quantum devices, the couplings can be controlled by various methods, as demonstrated in recent experiments~\cite{squbit_rev, BlaisRMP2021cQED, SiddiqiPRL2021, CampbellLaHayePRApplied2023, Chen:2014, FYanPRApplied2018, SandbergAPL2008}, and the ratio $\alpha$ can be well below $5\%$~\cite{GeorgopoulosPRA2021, error_range}.
In Fig.~\ref{fig5}(a), the fidelity $F(t)$ is plotted vs $t/T$ for a single sample of errors with $\alpha=0.05$ and $T=0.5\pi$. The fidelity exhibits small fluctuations over the time $t$ due to the presence of the errors. In Fig.~\ref{fig5}(b), the fidelity $F(T)$ averaged over $100$ samples of errors is plotted vs the ratio $\alpha$. As $\alpha$ increases to $15\%$, the fidelity only shows a mild decrease, demonstrating that control errors will not strongly affect the system dynamics in practical systems.

\emph{Decoherence.} Environmental noise induces decoherence in the qubits and the cavity modes, potentially impacting the fidelity of the prepared quantum states. To analyze the effects of decoherence, we utilize a master equation approach with
\begin{equation}
    \frac{d\rho}{dt} = -i\left [H_{\rm tot}, \rho\right] + \sum_j \left(\gamma_j \mathcal{L}_{qj}+ \kappa_j\mathcal{L}_{aj}\right) \rho,
\end{equation}
where $\mathcal{L}_{qj} = \frac{1}{2}(2\sigma_{j-}\rho\sigma_{j+}- \rho\sigma_{j+}\sigma_{j-}-\sigma_{j+}\sigma_{j-}\rho)$ is the Liouvillian operator for the qubit on site $j$ with damping rate $\gamma_j$, $\mathcal{L}_{aj}=\frac{1}{2} (2a_j \rho a_j^\dagger- \rho a_j^\dagger a_j -a_j^\dagger a_j \rho)$ is the Liouvillian operator for the cavity mode on site $j$ with damping rate $\kappa_j$, and $H_{\rm tot}$ is the total Hamiltonian of the system. 
We choose $\gamma_{1,2}=\gamma$ and $\kappa_{1,2}=\kappa$ for simplicity of discussion. 
We also assume that the dimensionless coupling strength $g=1$ in our discussion corresponds to $g=2\pi\times 100$ MHz in superconducting systems~\cite{KCaiNpj2021, ParajuliSciRep2023}. 
The measured decoherence time of $\sim 100 {\rm \mu s}$ then corresponds to $\gamma=\frac{5}{\pi}\times10^{-5}$ in dimensionless units. When the cavity frequency is $\omega_0=2\pi\times 5$ GHz and the quality factor is $Q=10^6$, the cavity decay rate is $\kappa=5\times 10^{-5}$ in dimensionless units.

In Fig.~\ref{fig5}(c), we plot the infidelity $1-F(T)$ vs the total evolution time $T$ for $\kappa=5\times 10^{-5},\,10^{-4}$ and $5\times 10^{-4}$, respectively, with $\gamma=\frac{5}{\pi}\times 10^{-5}$. It can be seen that for a short evolution time $T=0.5\pi$, the infidelity under only the adiabatic Hamiltonian is much larger than the infidelity with the CD driving $H_{1}^{\prime}$, indicating that the infidelity is dominated by diabatic transitions. While for a long evolution time $T=5.5\pi$, the infidelities without and with the CD driving become comparable, dominated by the decoherence effect.
In Fig.~\ref{fig5}(d), we plot $1-F(T)$ vs the cavity damping rate $\kappa$ for $T=0.5\pi,\,3\pi$ and $5.5\pi$, respectively. Here the infidelity increases slightly with $\kappa$ for a short evolution time $T=0.5\pi$, even when considering a decoherence rate much higher than the experimental value. In contrast, for a long evolution time such as $T=5.5\pi$, the infidelity increases significantly with $\kappa$, indicating the dominance of the decoherence effect. This result clearly shows that by applying the implementable CD driving $H_{1}^{\prime}$, we can greatly reduce the decoherence effect by choosing a short evolution time and achieve high-fidelity quantum states. This finding confirms the effectiveness of the CD approach in mitigating the impact of decoherence. Meanwhile, as discussed in Sec.~\ref{ssec:S2E1}, the implementable CD driving $H_{1}^{\prime}$ generates the same system dynamics as the CD driving $H_{1}$ when the initial state is $v_{1}$. The effect of decoherence and circuit errors when $H_{1}^{\prime}$ is applied is hence the same as that when $H_{1}$ is applied. Using $H_{1}^{\prime}$ will not be more robust to noise and control errors than using $H_{1}$.

\begin{figure}[t]
\includegraphics[clip,width=8.5cm]{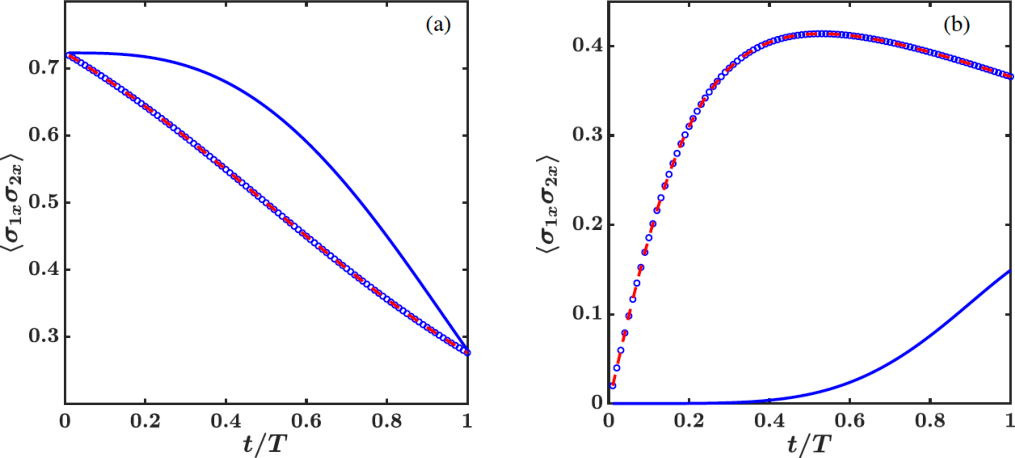}
\caption{(Color online) Operator average $\langle\sigma_{1x}\sigma_{2x}\rangle$ vs $t/T$ for $T=0.5\pi$. (a) Two sites one excitation with linear ramping of $J$ and parameters given in Fig.~\ref{fig3} in dimensionless units.  (b) Two sites two excitations with linear ramping of $J$ and parameters given in Fig.~\ref{fig4} in dimensionless units.
In (a) and (b), blue solid line (blue circles): under the adiabatic Hamiltonian $H_{\rm r}$ only (with the CD driving $H_1^\prime$ applied); and  
red dashed line: instantaneous ground state average of $H_{\rm r}$.}
\label{fig6} 
\end{figure} 
\emph{Measurement.} The performance of the CD approach can be characterized by conducting measurement on the qubit operator $\sigma_{1x}\sigma_{2x}$. In the case of one excitation in a two-site lattice, the ground state at the initial parameters $g=1$, $J=0$, and $\Delta=1$ is $\frac{1}{2}(\vert 1,-\rangle_1 \vert 0,g\rangle_2+ \vert 0,g\rangle_1 \vert 1,-\rangle_2)$, where $\vert 1,- \rangle$ is the lower-polariton state with one excitation, and the excitation occupies the sites in the form of a quantum superposition with the operator average $\langle\sigma_{1x}\sigma_{2x}\rangle=0.7236$. When the hopping rate becomes $J_f=2$ at time $T$, the operator average of the ground state decreases to $\langle\sigma_{1x}\sigma_{2x}\rangle=0.2764$, as shown in Fig.~\ref{fig6}(a). It is noticeable that the operator average for the adiabatic evolution under $H_{\rm r}$ can significantly deviate from that of the ground state, while the operator average with the CD driving applied remains the same as that of the ground state. 
Similarly, in the two-excitation case, the ground state at $g=1$, $J=0$, and $\Delta=0$ is $v_7=\vert 1,- \rangle_1 \vert 1,-\rangle_2$, where each site is occupied by one excitation in the lower-polariton state with the operator average $\langle\sigma_{1x}\sigma_{2x}\rangle=0$.  When the hopping rate becomes $J_f=1$, the operator average increases to $\langle\sigma_{1x}\sigma_{2x}\rangle=0.3659$. During the evolution, the operator average for the adiabatic process without the CD driving shows a large discrepancy with the operator average in the CD approach, as shown in Fig.~\ref{fig6}(b). Hence, the effectiveness of the CD approach can be verified by measuring the operator average.

\section{Conclusions \label{sec:conlusions}}
To conclude, we have studied a quantum shortcut to adiabaticity using the counter-diabatic driving approach in a finite-sized JC lattice, leveraging the symmetry inherent in its eigenstates. Our analysis reveals that for a two-site lattice with one excitation, the CD driving can be simplified to a realizable form that only consists of local qubit-cavity couplings. For more complex scenarios, such as a two-site lattice with two excitations or a three-site lattice with one excitation, a simpler CD Hamiltonian can still be obtained, albeit involving nonlocal terms. Our numerical simulations demonstrate that control errors and environmental noise have negligible effects on our scheme in practical systems and our scheme can be characterized through measurements on the qubits. Overall, our findings offer a promising avenue for achieving high-fidelity quantum state preparation and hold potential for applications across various systems exhibiting similar symmetries.

\section*{Acknowledgements}
This work is supported by the NSF Award No. 2037987 and the UC-MRPI Program (Grant ID M23PL5936). A.G. is also supported by the Sandbox AQ Fellowship. 

\begin{appendix}

\section{Relation between $H_{1}$ and $H_{1}^{\prime}$}
\begin{figure}[t]
\includegraphics[clip, width=8.5cm]{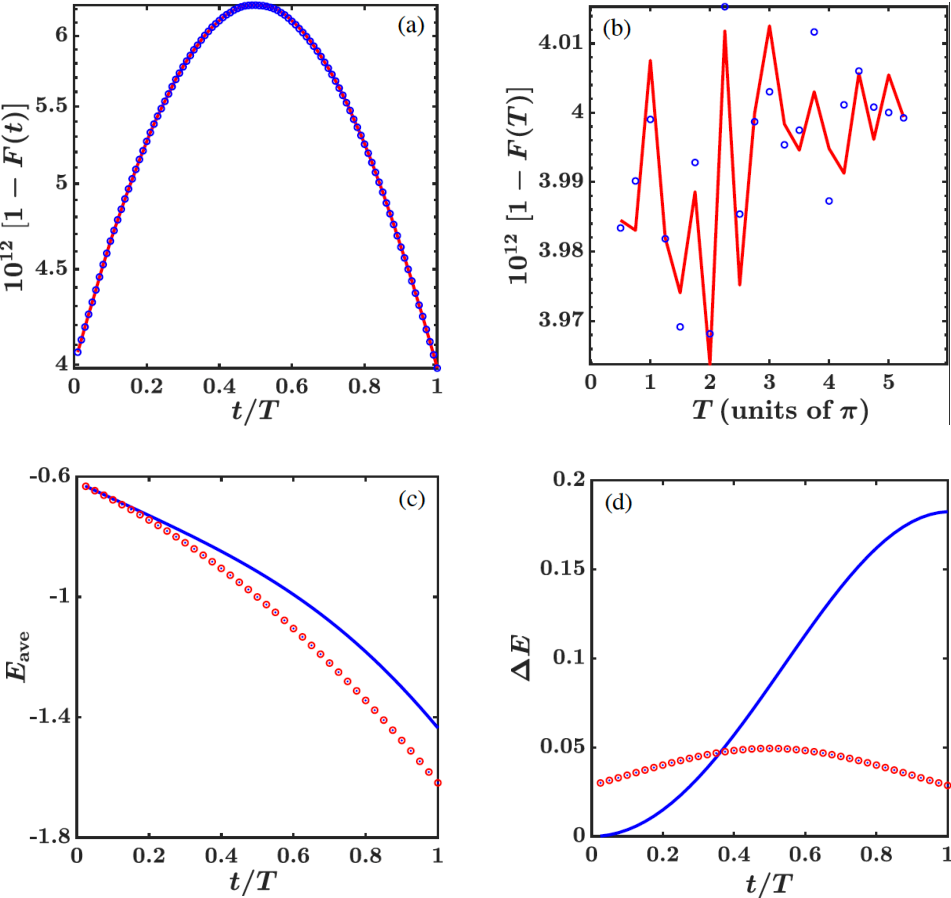}
\caption{(Color online) (a) Infidelity $1-F(t)$ vs $t/T$ for total evolution time $T=0.5 \pi$. (b) Infidelity $1-F(T)$ vs $T$. In (a) and (b), red solid line (blue circles): under the adiabatic Hamiltonian $H_{\rm r}$ with the CD driving $H_{1}$ ($H_1^\prime$) applied. 
(c) Average energy $E_{\rm ave}$ vs $t/T$. (d) Energy cost $\Delta E$ vs $t/T$. In (c) and (d), blue solid line: under $H_{\rm r}$ only; red circles: with $H_{1}$ applied; and blue dots: with $H_{1}^{\prime}$ applied. 
All the plots are for two sites one excitation with linear ramping of $J$, and the parameters are given in Fig.~\ref{fig3} in dimensionless unit.}
\label{fig7}
\end{figure}

In Sec.~\ref{ssec:S2E1}, we give an analysis that the system dynamics under the CD driving $H_{1}^{\prime}$ will be exactly the same as that of the CD driving $H_{1}$, when the initial state is $v_{1}$. In Fig.~\ref{fig3}(a, b), we plot the infidelity for a two-site lattice with one excitation under the CD driving $H_{1}^{\prime}$, where the infidelity remains zero up to a small numerical error below $10^{-10}$. In Fig.~\ref{fig7}(a, b), we plot the infidelity of this system under both $H_{1}$ and $H_{1}^{\prime}$ for a linear ramping of the hopping rate $J$ using the same parameters as given in Fig.~\ref{fig3}(a, b). It can be clearly seen that the infidelities for both $H_{1}$ and $H_{1}^{\prime}$ remain zero up to the numerical error, which confirms our analysis.

Furthermore, we compare the effect of the CD drivings $H_{1}$ and $H_{1}^{\prime}$ by numerically calculating two quantities:  (1) the average energy of the system $E_{\rm ave} = \langle v(t) | H_{\rm tot}(t) |v(t)\rangle $ at time $t$ with $H_{\rm tot}$ being the total Hamiltonian and $v(t)$ being the instantaneous wavefunction, and (2) the energy cost $\Delta E = E_{\rm ave} - E_{G}$ with $E_{G}$ being the instantaneous ground state energy at time $t$ following the definition in \cite{AbahPRE2019}.
In Fig.~\ref{fig7}(c), we plot $E_{\rm ave}$ for linear ramping of the hopping rate $J$ on the initial state $\vert v_{1} \rangle $ under the adiabatic Hamiltonian only, with the CD driving $H_{1}$ applied, and with $H_{1}^{\prime}$ applied, respectively. The average energy for the adiabatic Hamiltonian  $H_{\rm tot} = H_{\rm r}$ is higher than that with the CD driving $H_{1}$ or $H_{1}^{\prime}$, indicating nonzero diabatic transition to the excited states, while the average energy with the CD driving $H_{1}$ is the same as that with $H_{1}^{\prime}$, confirming that these two CD Hamiltonians give the same dynamics. 
In Fig.~\ref{fig7}(d), we plot the energy cost $\Delta E$ vs the time $t/T$. For the adiabatic process under the Hamiltonian $H_{\rm tot}=H_{\rm r}$, $\Delta E=0$ at time $t=0$, because the initial state is the ground state of the instantaneous Hamiltonian. During the evolution, $\Delta E$ continuously increases due to diabatic transition to the excited state. Under the CD driving $H_{1}$ or $H_{1}^{\prime}$, the energy cost is nonzero. This is because the state $v(t)$ at an arbitrary time $t$ is the instantaneous ground state of $H_{\rm r}$ but not the ground state of the total Hamiltonian with the CD driving $H_{1}$ or $H_{1}^{\prime}$ being nonzero at any time $t$ including the initial and the final times.

\section{CD Driving for Two Excitations}
For a two-site lattice with two polariton excitations, the adiabatic Hamiltonian is given in Eq.~(\ref{eq:Htr2}). Solving this Hamiltonian for $\Delta=0$, we can find its eigenvalues and eigenvectors. The eigenvalues are given in (\ref{eq:E1_8}), and the eigenvectors are
\begin{subequations}
\begin{align}
\vert v_{1}\rangle &=\left( 0, -J A_{1}, 0, -J A_{1}, - g A_{1}, 0,0,1\right),  \nonumber  \\
\vert v_{2}\rangle &= \left( A_{2}, 0, -A_{2},0,0,-1,1,0\right),  \nonumber \\
\vert v_{3,4}\rangle &=\left( -\frac{1}{A_{2}}, \pm \frac{1}{A_{3}},  \frac{1}{A_{2}},\mp \frac{1}{A_{3}},0,-1,1,0\right),  \nonumber  \\
\vert v_{5,6}\rangle &=\left(\pm X_{-}, Y_{-}, \pm X_{-}, Y_{-}, Z_{-}, \pm M_{-}, \pm M_{-}, 1\right),  \nonumber  \\
\vert v_{7,8}\rangle &=\left( \mp X_{+}, Y_{+}, \mp X_{+}, Y_{+}, Z_{+}, \mp M_{+}, \mp M_{+}, 1\right),  \nonumber 
\end{align}
\label{eq:v1_8}
\end{subequations}
where we define
\begin{widetext}
\begin{subequations}
\begin{align}
A_{1} & = \frac{g}{g^{2}- J^{2}},\,\, A_{2} = -\frac{J}{\sqrt{2}g},\,\, A_{3} = -\frac{J}{\sqrt{2g^{2}+J^{2}}},\,\, 
A_{4} = \sqrt{4 g^4 + 60 g^2 J^2 + 9 J^4},   \nonumber  \\
X_{\pm}&=-\frac{24J}{g^{2}}\frac{[\pm 2g^{6}+g^{4}(\pm 19J^{2}+A_{4})+J^{4}(\pm 3J^{2}+A_{4})+g^{2}(\pm 19J^{4}+3J^{2}A_{4})]}{\sqrt{6g^{2}+5J^{2}\pm A_{4}}(\pm 2g^{2} \pm 9J^{2}+A_{4})(\pm 2g^{2} \pm 3J^{2}+A_{4})}, \nonumber \\
Y_{\pm}&=-\frac{8J}{g}\frac{[16g^{6}+g^{4}(166J^{2}\pm 8A_{4}) \pm 12J^{4}( \pm 3J^{2}+A_{4})+g^{2}(201J^{4} \pm 29J^{2}A_{4})]}{(2g^{2}+3J^{2}\pm A_{4})(6g^{2}+5J^{2} \pm A_{4})(2g^{2}+9J^{2} \pm A_{4})},  \nonumber\\
Z_{\pm }&=\frac{2}{g^{2}}\frac{[8g^{6}+g^{4}(90J^{2}\pm 4A_{4})\pm 8J^{4}(3J^{2}+A_{4})+g^{2}(125J^{4}\pm 17J^{2}A_{4})]}{(\pm 2g^{2}\pm 9J^{2}+A_{4})(\pm 6g^{2} \pm 5J^{2}+A_{4})},  \nonumber\\
M_{\pm }&=\frac{2\sqrt{2}}{g}\frac{[\pm 2g^{4}+g^{2}(\pm 11J^{2}+A_{4})+J^{2}(\pm 3J^{2}+A_{4})]}{(2g^{2}+3J^{2} \pm A_{4})\sqrt{6g^{2}+5J^{2}\pm A_{4}}}.  \nonumber
\end{align}
\label{eq:A_4_XYZM}
\end{subequations}
\end{widetext}
Note that the eigenvectors given above have not been normalized. We find that the eigenvectors can be grouped into two subsets. The first subset is composed of the antisymmetric eigenvectors $\{v_{2}, v_{3}, v_{4}\}$, and the second subset is composed of the symmetric eigenvectors $\{v_{1}, v_{5}, v_{6}, v_{7}, v_{8}\}$.

Using the expressions of the eigenvalues, eigenvectors, and Eq.~(\ref{eq:H1}), we can derive the CD driving $H_{1}$ for this system, but we cannot write $H_{1}$ into an analytical form. Neither can we find an implementable CD driving as we did in Sec.~\ref{ssec:S2E1} for the case of one excitation. 
As varying the adiabatic Hamiltonian $H_{\rm r}$ only induces diabatic transitions between eigenstates of the same symmetry, the CD driving only contains nonzero matrix elements between eigenstates in the same subset.  We find that when considering only the matrix elements for the first subset $\{v_{2}, v_{3}, v_{4}\}$, the CD Hamiltonian can be written in terms of the qubit and cavity operators as
\begin{widetext}
\begin{equation}
H_{1}^{\prime} = i\frac{g}{2(J^{2}+2g^{2})}\frac{\partial J}{\partial t}\left[(a_{1}^{\dagger})^{2}\sigma_{1-}a_{2}-(a_{1}^{\dagger})^{2}a_{1}\sigma_{2-}-(a_{2}^{\dagger})^{2}\sigma_{1-} a_{2}+(a_{2}^{\dagger})^{2}a_{1}\sigma_{2-}\right]+h.c.  \label{eq:H1deriveS2E2}
\end{equation}
\end{widetext}
Defining $A_2 = a_{1}^2 - a_{2}^2$ and $C_2= a_{2}\sigma_{1-}-a_{1}\sigma_{2-}$, it can be shown that Eq.~(\ref{eq:H1deriveS2E2}) leads to Eq.~(\ref{eq:H1S2E2}) in Sec.~\ref{ssec:S2E2}. 
On the other hand, we cannot find an analytical expression for the CD driving for the second subset that contains the ground state $v_{7}$. 

\end{appendix}


\begin{thebibliography}{10}

\bibitem{Albash2018}
T. Albash and D.~A. Lidar,
\newblock Adiabatic quantum computation,
\newblock {Rev. Mod. Phys.} \textbf{90}, 015002 (2018).

\bibitem{FarhiScience2001}
E. Farhi, J. Goldstone, S. Gutmann, J. Lapan, A. Lundgren, and D. Preda,
\newblock A quantum adiabatic evolution algorithm applied to random instances of an NP-complete problem,
\newblock {Science} \textbf{292}, 472 (2001).

\bibitem{Preskill_2018}
J. Preskill,
\newblock Quantum computing in the NISQ era and beyond,
\newblock {Quantum} \textbf{2}, 79 (2018).

\bibitem{Bharti_2021}
K. Bharti, A. Cervera-Lierta, T.~H. Kyaw, T. Haug, S.
  Alperin-Lea, A. Anand, M. Degroote, H. Heimonen, J.~S.
  Kottmann, T. Menke, W.-K Mok, S. Sim, L.-C Kwek, and A.
  Aspuru-Guzik,
\newblock Noisy intermediate-scale quantum algorithms,
\newblock {\em Rev. Mod. Phys.} \textbf{94}, 015004 (2022).

\bibitem{Guery-Odelin2019}
D.~Gu\'ery-Odelin, A.~Ruschhaupt, A.~Kiely, E.~Torrontegui, S.~Mart\'{\i}nez-Garaot, and J.~G. Muga,
\newblock Shortcuts to adiabaticity: Concepts, methods, and applications,
\newblock {Rev. Mod. Phys.} \textbf{91}, 045001 (2019).

\bibitem{Kolodrubetz2017}
M. Kolodrubetz, D. Sels, P.Mehta, and A. Polkovnikov,
\newblock Geometry and non-adiabatic response in quantum and classical systems,
\newblock {Phys. Rep.} \textbf{697}, 1 (2017).

\bibitem{UnanyanOptCommon1997}
R. G. Unanyan, L. P. Yatsenko, K. Bergmann, and B.W. Shore,  
\newblock Laser-induced adiabatic atomic reorientation with control of diabatic losses,
\newblock {Opt. Commun.} \textbf{139}, 48 (1997).

\bibitem{EmmanouilidouPRL2000}
A. Emmanouilidou, X.-G. Zhao, P. Ao, and Q. Niu,
\newblock Steering an Eigenstate to a Destination,
\newblock {Phys. Rev. Lett.} \textbf{85}, 1626 (2000).

\bibitem{Berry2009}
M.~V. Berry,
\newblock Transitionless quantum driving,
\newblock {J. Phys. A: Math. Theor.},
  \textbf{42}, 365303 (2009).

\bibitem{Demirplak2008}
M. Demirplak and S.~A. Rice,
\newblock {On the consistency, extremal, and global properties of
  counterdiabatic fields},
\newblock {J. Chem. Phys.} \textbf{129}, 154111 (2008).

\bibitem{Chen:PhysRevLett:2010:105}
X.~Chen, I.~Lizuain, A.~Ruschhaupt, D.~Gu\'ery-Odelin, and J.~G. Muga,
\newblock Shortcut to adiabatic passage in two- and three-level atoms,
\newblock {Phys. Rev. Lett.} \textbf{105}, 123003 (2010).

\bibitem{delCampoPRL2012}
A.~del Campo, M.~M.~Rams, and W.~H. Zurek,
\newblock Assisted finite-rate adiabatic passage across a quantum critical point: Exact solution for the quantum ising model, 
\newblock {Phys. Rev. Lett.} \textbf{109}, 115703 (2012).

\bibitem{Damski2014}
B. Damski,
\newblock Counterdiabatic driving of the quantum ising model,
\newblock {J. Stat. Mech.: Theory Exp.} \textbf{2014}(12), P12019 (2014).

\bibitem{Chen:PhysRevLett:2010:104}
X.~Chen, A.~Ruschhaupt, S.~Schmidt, A.~del Campo, D.~Gu\'ery-Odelin, and J.~G.~Muga,
\newblock Fast optimal frictionless atom cooling in harmonic traps: Shortcut to adiabaticity, 
\newblock {Phys. Rev. Lett.} \textbf{104}, 063002 (2010).

\bibitem{Jing2013}
J. Jing, L.-A Wu, M.~S. Sarandy, and J.~G. Muga,
\newblock Inverse engineering control in open quantum systems,
\newblock {Phys. Rev. A} \textbf{88}, 053422 (2013).

\bibitem{Motzoi2009}
F.~Motzoi, J.~M. Gambetta, P.~Rebentrost, and F.~K. Wilhelm,
\newblock Simple pulses for elimination of leakage in weakly nonlinear qubits,
\newblock {Phys. Rev. Lett.} \textbf{103}, 110501 (2009).

\bibitem{Baksic2016}
A. Baksic, H. Ribeiro, and A.~A. Clerk,
\newblock Speeding up adiabatic quantum state transfer by using dressed states,
\newblock {Phys. Rev. Lett.} \textbf{116}, 230503 (2016).

\bibitem{Wang2018}
T.~Wang, Z. Zhang, L. Xiang, Z. Jia, P. Duan, W.~Cai, Z.~Gong, Z.~Zong, M.~Wu, J.~Wu, L.~Sun, Y.~Yin, and G.~Guo,
\newblock The experimental realization of high-fidelity
  `shortcut-to-adiabaticity' quantum gates in a superconducting xmon qubit,
\newblock {New J. Phys.} \textbf{20}, 065003 (2018).

\bibitem{Vepsalainen2018}
A.~Veps\"{a}l\"{a}inen, S.~Danilin, and G.S.~Paraoanu, 
\newblock Optimal superadiabatic population transfer and gates by dynamical phase corrections,
\newblock {Quantum Sci. and Technol.} \textbf{3}, 024006 (2018).

\bibitem{Theis2016}
L.~S. Theis, F.~Motzoi, and F.~K. Wilhelm,
\newblock Simultaneous gates in frequency-crowded multilevel systems using
  fast, robust, analytic control shapes,
\newblock {Phys. Rev. A} \textbf{93}, 012324 (2016).

\bibitem{Martinis2014}
J.~M. Martinis and M.~R. Geller,
\newblock Fast adiabatic qubit gates using only ${\ensuremath{\sigma}}_{z}$
  control,
\newblock {Phys. Rev. A} \textbf{90}, 022307 (2014).

\bibitem{Yan2019}
T. Yan, B-J. Liu, K. Xu, C. Song, S. Liu, Z. Zhang, H.
  Deng, Z. Yan, H. Rong, K. Huang, M.-H. Yung, Y. Chen,
  and D. Yu,
\newblock Experimental realization of nonadiabatic shortcut to non-abelian
  geometric gates,
\newblock {Phys. Rev. Lett.} \textbf{122}, 080501 (2019).

\bibitem{Liang2016}
Z.-T. Liang, X. Yue, Q. Lv, Y.-X. Du, W. Huang, H. Yan,
  and S.-L. Zhu,
\newblock Proposal for implementing universal superadiabatic geometric quantum
  gates in nitrogen-vacancy centers,
\newblock {Phys. Rev. A} \textbf{93}, 040305(R) (2016).

\bibitem{Du2016}
Y.-X. Du, Z.-T. Liang, Y.-C. Li, X.-X. Yue, Q.-X. Lv, W.~Huang, X.~Chen, H. Yan, and S.-L. Zhu,
\newblock Experimental realization of stimulated raman shortcut-to-adiabatic
  passage with cold atoms,
\newblock {Nat. Commun.} \textbf{7}, 12479 (2016).

\bibitem{Zhou2017}
B.B. Zhou, A. Baksic, H. Ribeiro, C.G. Yale,
  F.J. Heremans, P.C. Jerger, A. Auer, G. Burkard, A.A. Clerk, and D.D. Awschalom,
\newblock Accelerated quantum control using superadiabatic dynamics in a
  solid-state lambda system,
\newblock {Nat. Phys.} \textbf{13}, 330 (2017).

\bibitem{Masuda2018}
S. Masuda, K.~Y. Tan, and M. Nakahara,
\newblock Spin-selective electron transfer in a quantum dot array,
\newblock {Phys. Rev. B} \textbf{97}, 045418 (2018).

\bibitem{Bason:NatPhys:2011}
M.~G. Bason, M. Viteau, N. Malossi, P. Huillery, E. Arimondo, D. Ciampini, R. Fazio, V. Giovannetti, R. Mannella, and O. Morsch,
\newblock High-fidelity quantum driving,
\newblock {Nat. Phys.} \textbf{8}, 147 (2012).

\bibitem{Hatomura2018}
T. Hatomura,
\newblock Shortcuts to adiabatic cat-state generation in bosonic Josephson junctions,
\newblock {New J. Phys.} \textbf{20}, 015010 (2018).

\bibitem{Stefanatos2018}
D. Stefanatos and E. Paspalakis,
\newblock Maximizing entanglement in bosonic josephson junctions using
  shortcuts to adiabaticity and optimal control.
\newblock {New J. Phys.} \textbf{20}, 055009 (2018).

\bibitem{Zhang2015}
J.~Zhang, T.~H. Kyaw, D.~M. Tong, E. Sj\"{o}qvist, and L.-C. Kwek,
\newblock Fast non-abelian geometric gates via transitionless quantum driving,
\newblock {Sci. Rep.} \textbf{5}, 18414 (2015).

\bibitem{delCampoPRL2013}
A. del Campo,
\newblock Shortcuts to adiabaticity by counterdiabatic driving,
\newblock {Phys. Rev. Lett.} \textbf{111}, 100502 (2013).

\bibitem{DeffnerPRX2014}
S. Deffner, C. Jarzynski, and A. del Campo,
\newblock Classical and Quantum Shortcuts to Adiabaticity for Scale-Invariant Driving,
\newblock {Phys. Rev. X} \textbf{4}, 021013 (2014).

\bibitem{Sels2017}
D. Sels and A. Polkovnikov,
\newblock Minimizing irreversible losses in quantum systems by local
  counterdiabatic driving,
\newblock {Proc. Natl. Acad. Sci.} \textbf{114}, E3909 (2017).

\bibitem{Takahashi2017}
K. Takahashi,
\newblock Shortcuts to adiabaticity for quantum annealing,
\newblock {Phys. Rev. A} \textbf{95}, 012309 (2017).

\bibitem{Opatrny2014}
T. Opatrn\'{y} and K. M{\o}lmer,
\newblock Partial suppression of nonadiabatic transitions,
\newblock {New J. Phys.} \textbf{16}, 015025 (2014).

\bibitem{BoyersPRA2019}
E. Boyers, M. Pandey, D. K. Campbell, A.~Polkovnikov, D. Sels, and A. O. Sushkov,
\newblock Floquet-engineered quantum state manipulation in a noisy qubit,
\newblock {Phys. Rev. A} \textbf{100}, 012341 (2019). 

\bibitem{ClaeysPRL2019}
P. W. Claeys, M. Pandey, D. Sels, and A.~Polkovnikov, 
\newblock Floquet-Engineering Counterdiabatic Protocols in Quantum Many-Body Systems,
\newblock {Phys. Rev. Lett.} \textbf{123}, 090602 (2019). 

\bibitem{Hegade2022L}
N.~N. Hegade, X.~Chen, and E. Solano,
\newblock Digitized counterdiabatic quantum optimization,
\newblock {Phys. Rev. Res.} \textbf{4}, L042030 (2022).

\bibitem{KeeverPRXQuantum2024}
C. M. Keever and M. Lubasch,
\newblock Towards Adiabatic Quantum Computing Using Compressed Quantum Circuits,
\newblock {PRX Quantum} \textbf{5}, 020362 (2024).

\bibitem{CepaitePRXQuantum2023}
I. \v{C}epait\.{e}, A. Polkovnikov, A. J. Daley, and C.~W.~Duncan, 
\newblock Counterdiabatic Optimized Local Driving,
\newblock {PRX Quantum} \textbf{4}, 010312 (2023).

\bibitem{Hartmann:2006}
M.~J. Hartmann, F. G. S.~L. Brand\~{a}o, and M.~B. Plenio,
\newblock Strongly interacting polaritons in coupled arrays of cavities,
\newblock {Nat. Phys.} \textbf{2}, 849 (2006).

\bibitem{Greentree:2006}
A.~D. Greentree, C. Tahan, J.~H. Cole, and L. C.~L. Hollenberg,
\newblock Quantum phase transitions of light,
\newblock {Nat. Phys.} \textbf{2}, 856 (2006).

\bibitem{Angelakis:2007}
D.~G. Angelakis, M.~F. Santos, and S. Bose,
\newblock Photon-blockade-induced mott transitions and $xy$ spin models in
  coupled cavity arrays,
\newblock {Phys. Rev. A} \textbf{76}, 031805(R) (2007).

\bibitem{2007RossiniPRL_JC}
D. Rossini and R. Fazio,
\newblock Mott-insulating and glassy phases of polaritons in 1d arrays of
  coupled cavities,
\newblock {Phys. Rev. Lett.} \textbf{99}, 186401 (2007).

\bibitem{2008NeilPra_BH}
N. Na, S. Utsunomiya, L.~Tian, and Y. Yamamoto,
\newblock Strongly correlated polaritons in a two-dimensional array of photonic
  crystal microcavities,
\newblock {Phys. Rev. A} \textbf{77}, 031803(R) (2008).

\bibitem{2009KochPra_QS}
J. Koch and K. Le~Hur,
\newblock Superfluid--mott-insulator transition of light in the jaynes-cummings lattice,
\newblock {Phys. Rev. A} \textbf{80}, 023811 (2009).

\bibitem{Seo2015:1}
K. Seo and L. Tian,
\newblock Quantum phase transition in a multiconnected superconducting
  jaynes-cummings lattice,
\newblock {Phys. Rev. B} \textbf{91}, 195439 (2015).

\bibitem{Xue2017}
J. Xue, K. Seo, L. Tian, and T. Xiang,
\newblock Quantum phase transition in a multiconnected jaynes-cummings lattice,
\newblock {Phys. Rev. B} \textbf{96}, 174502 (2017).

\bibitem{Noh2017Review}
C. Noh and D.~G.~Angelakis,
\newblock Quantum simulations and many-body physics with light,
\newblock {Rep. Prog. Phys.} \textbf{80}, 016401 (2016).

\bibitem{2012HouckNP_JCQS}
A.~A. Houck, H.~E. T\"{u}reci, and J. Koch,
\newblock On-chip quantum simulation with superconducting circuits,
\newblock {Nat. Phys.} \textbf{8}, 292 (2012).

\bibitem{Hoffman:2011}
A.~J. Hoffman, S.~J. Srinivasan, S.~Schmidt, L.~Spietz, J.~Aumentado, H.~E.
  T\"ureci, and A.~A. Houck,
\newblock Dispersive photon blockade in a superconducting circuit,
\newblock {Phys. Rev. Lett.} \textbf{107}, 053602 (2011).

\bibitem{Sala2015Nanophotonics}
V.~G. Sala, D.~D. Solnyshkov, I.~Carusotto, T.~Jacqmin, A.~Lema\^{\i}tre,
  H.~Ter\ifmmode~\mbox{\c{c}}\else \c{c}\fi{}as, A.~Nalitov, M.~Abbarchi,
  E.~Galopin, I.~Sagnes, J.~Bloch, G.~Malpuech, and A.~Amo,
\newblock Spin-orbit coupling for photons and polaritons in microstructures,
\newblock {Phys. Rev. X} \textbf{5}, 011034 (2015).

\bibitem{Lepert2011Atom}
G~Lepert, M~Trupke, M~J Hartmann, M~B Plenio, and E~A Hinds,
\newblock Arrays of waveguide-coupled optical cavities that interact strongly
  with atoms,
\newblock {New J. Phys.} \textbf{13}, 113002 (2011).

\bibitem{Ivanov2009Ion}
P.~A. Ivanov, S.~S. Ivanov, N.~V. Vitanov, A.~Mering, M.~Fleischhauer, and K.~Singer,
\newblock Simulation of a quantum phase transition of polaritons with trapped ions,
\newblock {Phys. Rev. A} \textbf{80}, 060301(R) (2009).

\bibitem{Toyoda2013Ion}
K. Toyoda, Y. Matsuno, A. Noguchi, S. Haze, and S. Urabe,
\newblock Experimental realization of a quantum phase transition of polaritonic
  excitations,
\newblock {Phys. Rev. Lett.} \textbf{111}, 160501 (2013).

\bibitem{Debnath2018Ion}
S.~Debnath, N.~M. Linke, S.-T. Wang, C.~Figgatt, K.~A. Landsman, L.-M. Duan, and C.~Monroe,
\newblock Observation of hopping and blockade of bosons in a trapped ion spin  chain,
\newblock {Phys. Rev. Lett.} \textbf{120}, 073001 (2018).

\bibitem{BWLi2022Ion}
B.-W. Li, Q.-X. Mei, Y.-K. Wu, M.-L. Cai, Y.~Wang, L.~Yao, Z.-C. Zhou, and L.-M. Duan,
\newblock Observation of non-markovian spin dynamics in a  Jaynes-Cummings-Hubbard model using a trapped-ion quantum simulator,
\newblock {Phys. Rev. Lett.} \textbf{129}, 140501 (2022).

\bibitem{HouckPRX2017}
M. Fitzpatrick, N.~M. Sundaresan, A. C.~Y. Li, J. Koch, and A.~A. Houck,
\newblock Observation of a dissipative phase transition in a one-dimensional
  circuit qed lattice,
\newblock {Phys. Rev. X} \textbf{7}, 011016 (2017).

\bibitem{KeelingPRL2012}
F. Nissen, S. Schmidt, M. Biondi, G. Blatter, H.~E.~T\"{u}reci, and J. Keeling,
\newblock Nonequilibrium dynamics of coupled qubit-cavity arrays,
\newblock {Phys. Rev. Lett.} \textbf{108}, 233603 (2012).

\bibitem{Larson2022Review}
J. Larson and T. Mavrogordatos,
\newblock \emph{The Jaynes-Cummings Model and Its Descendants: Modern research directions}, 
\newblock Part of IOP Series in Quantum Technology (2021), see also arXiv:2202.00330.

\bibitem{TianPRL2011}
Y. Hu and L. Tian,
\newblock Deterministic generation of entangled photons in superconducting resonator arrays,
\newblock {Phys. Rev. Lett.} \textbf{106}, 257002 (2011).

\bibitem{KCaiNpj2021}  
K. Cai, P. Parajuli, G. Long, C.~W. Wong, and L. Tian,
\newblock Robust preparation of many-body ground states in Jaynes-Cummings lattices,
\newblock {Npj Quantum Inf.} \textbf{7}, 96 (2021).

\bibitem{ParajuliSciRep2023}
P. Parajuli, A. Govindarajan, and L. Tian,
\newblock State preparation in a Jaynes-Cummings lattice with quantum optimal control,
\newblock {Sci. Rep.} \textbf{13}, 19924 (2023).

\bibitem{FunoNoriPRL2021}
K. Funo, N. Lambert, and F. Nori, 
\newblock General Bound on the Performance of Counter-Diabatic Driving Acting on Dissipative Spin Systems,
\newblock {Phys. Rev. Lett.} \textbf{127}, 150401 (2021).

\bibitem{AlipourQuantum2020}
S. Alipour, A. Chenu, A. T. Rezakhani, and A. del Campo, 
\newblock Shortcuts to Adiabaticity in Driven Open Quantum Systems: Balanced Gain and Loss and Non-Markovian Evolution,
\newblock {Quantum} \textbf{4}, 336 (2020).


\bibitem{squbit_rev}
P.~Krantz, M.~Kjaergaard, F.~Yan, T.~P. Orlando, S.~Gustavsson, and W.~D.~Oliver,
\newblock {A quantum engineer's guide to superconducting qubits},
\newblock {Appl. Phys. Rev.} \textbf{6}, 021318 (2019). 

\bibitem{BlaisRMP2021cQED}
A. Blais, A.~L. Grimsmo, S.~M. Girvin, and A. Wallraff,
\newblock Circuit quantum electrodynamics,
\newblock {Rev. Mod. Phys.} \textbf{93}, 025005 (2021).

\bibitem{SiddiqiPRL2021}
B.~K. Mitchell, R.~K. Naik, A. Morvan, A. Hashim, J.~M.~Kreikebaum, B.~Marinelli, W.~Lavrijsen, K.~Nowrouzi, D.~I. Santiago, and I.~Siddiqi,
\newblock Hardware-efficient microwave-activated tunable coupling between superconducting qubits,
\newblock {Phys. Rev. Lett.} \textbf{127}, 200502 (2021).

\bibitem{CampbellLaHayePRApplied2023}
D.~L. Campbell, A. Kamal, L. Ranzani, M. Senatore, and M.~D. LaHaye,
\newblock Modular tunable coupler for superconducting circuits,
\newblock {Phys. Rev. Appl.} \textbf{19}, 064043 (2023).

\bibitem{Chen:2014}
Y.~Chen, C.~Neill, P.~Roushan, N.~Leung, M.~Fang, R.~Barends, J.~Kelly,
  B.~Campbell, Z.~Chen, B.~Chiaro, A.~Dunsworth, E.~Jeffrey, A.~Megrant, J.~Y.
  Mutus, P.~J.~J. O'Malley, C.~M. Quintana, D.~Sank, A.~Vainsencher, J.~Wenner,
  T.~C. White, Michael~R. Geller, A.~N. Cleland, and J.~M. Martinis,
\newblock Qubit architecture with high coherence and fast tunable coupling,
\newblock {Phys. Rev. Lett.} \textbf{113}, 220502 (2014).

\bibitem{FYanPRApplied2018}
F. Yan, P. Krantz, Y. Sung, M. Kjaergaard, D.~L.~Campbell,
  T.~P.~Orlando, S. Gustavsson, and W.~D.~Oliver,
\newblock Tunable coupling scheme for implementing high-fidelity two-qubit
  gates,
\newblock {Phys. Rev. Appl.} \textbf{10}, 054062 (2018).

\bibitem{SandbergAPL2008}
M.~Sandberg, C.~M. Wilson, F.~Persson, T.~Bauch, G.~Johansson, V.~Shumeiko,
  T.~Duty, and P.~Delsing,
\newblock Tuning the field in a microwave resonator faster than the photon lifetime,
\newblock {Appl. Phys. Lett.} \textbf{92}, 203501 (2008).

\bibitem{GeorgopoulosPRA2021}
K. Georgopoulos ,C. Emary, and P. Zuliani, 
\newblock Modeling and simulating the noisy behavior of near-term quantum computers,
\newblock {Phys. Rev. A} \textbf{104}, 062432 (2021). 

\bibitem{error_range}
For simplicity of discussion, we assume that the standard deviations of the circuit errors are $\alpha J_{f}$ and $\alpha g$, respectively. Under this assumption, the error amplitude is on the scale of a given percentage of the maximal hopping rate $J_{f}$ or the coupling strength $g$, which is reasonable in practical devices. Meanwhile, for $J_{f}=2$, $g=1$, $\Delta=1$, and $T\ge 0.5\pi$ ($dJ/dt = J_{f}/T$), the strength of the CD driving $g^{\prime}$ is on the same order as the magnitudes of $g$ and $J_{f}$, and hence, will not bring in large error beyond our assumption.

\bibitem{AbahPRE2019}
O. Abah and M. Paternostro
\newblock Shortcut-to-adiabaticity Otto engine: A twist to finite-time thermodynamics, 
\newblock {Phys. Rev. E} \textbf{99}, 022110 (2019).

\end{thebibliography}

\end{document}